\documentclass[11pt,letterpaper]{article}

\usepackage{amsmath}
\usepackage{amsfonts}
\usepackage{amssymb}
\usepackage{graphicx}
\usepackage[hang,small,bf]{caption}

\DeclareRobustCommand{\SkipTocEntry}[4]{}

\def\beq{\begin{equation}}
\def\eeq{\end{equation}}
\def\bea{\begin{eqnarray}}
\def\eea{\end{eqnarray}}

\usepackage{epsfig}
\usepackage{url}
\usepackage[hcentering,vcentering,hscale=0.7255,vscale=0.83]{geometry}

\usepackage{natbib}
\input{journals.cls}

\begin{document}
%-----------------------------------------------------------------------------------------------

%-----------------------------------------------------------------------------------------------
% Prologue
%-----------------------------------------------------------------------------------------------

%\vspace{10mm}

%\begin{center}

% Title

%{\Large CMBPol Workshop Proceedings}
%\\[1.0cm]

% Authors

%{A. Lazarian$^{\rm a}$}
%\\[0.5cm]

% Affiliations

%{\small
%\textit{$^{\rm a}$University of Wisconsin-Madison}}

%~

%DRAFT - DO NOT CIRCULATE

%\end{center}

% Table of Contents

%\newpage
%\tableofcontents

%-----------------------------------------------------------------------------------------------
% Section - Polarized Dust Emission
%-----------------------------------------------------------------------------------------------

%\newpage
\section{Grain Alignment by Radiative Torques}
\label{s:intro}
\begin{center}
{A. Lazarian, University of Wisconsin-Madison}
\end{center}

Grain alignment has a reputation of being a very tough astrophysical problem of a very long standing.
Indeed, for a long time since the discovery of dust-induced starlight extinction polarization in 1949 (Hall 1949; Hiltner 1949) the mechanism of alignment was both enigmatic and illusive.  Works by great minds like Lyman Spitzer and Edward Purcell moved the field forward in terms of understanding the basic grain dynamics physics, but, nevertheless, grain alignment theory did not have predictive powers for a long time.

Several mechanisms were proposed and elaborated to  various degree (see Lazarian 2007 for a review), including the "textbook solution", namely, 
the paramagnetic Davis-Greenstein (1951) mechanism, which matured through intensive work since its introduction (e.g. Jones \& Spitzer 1967, Purcell 1979, Spitzer \& McGlynn 1979, Mathis 1986, Roberge et al. 1993, Lazarian 1997, Roberge \& Lazarian 1999). The mechanical stochastic alignment was pioneered by Gold (1951), who concluded that supersonic flows should align grains rotating thermally. Further advancement of the mechanical alignment mechanism (e.g. Lazarian 1994, 1995a) allowed one to extend the range of applicability of the mechanism, but left it as an auxiliary process, nevertheless. Mechanical regular alignment of helical grains discussed in Lazarian (1995b), Lazarian, Goodman, Myers (1997), Lazarian \& Hoang (2007ab) seems to be more promising as it can aligned grains within subsonic flows. In any case, currently the mechanism based on radiative torques looks as the most promising (see Andersson \& Potter 2007, Lazarian 2007, Whittet et al. 2008). Thus we focus our short contribution on this mechanism. 

The effect of alignment induced by radiative torques was discovered by  Dolginov \& Mytrophanov (1976). They considered a grain which exhibited a difference in the cross-section for right-handed and left-handed photons. They noticed that scattering of unpolarized light by such a grain resulted in its spin-up. However, they could not quantify the effect and therefore their pioneering work was mostly neglected for the next 20 years. The explosion of interest to the radiative torques we owe to Bruce Draine, who realized that the torques can be treated with the modified version of the DDSCAT
code by Draine \& Flatau (1994). Empirical studies in Draine (1996), Draine \& Weingartner (1996, 1997), Weingartner \& Draine (2003) demonstrated that the magnitude of torques is substantial for irregular shapes studied. After that it became impossible to ignore the radiative torque alignment. Later, the spin-up of grains by radiative torques was demonstrated in laboratory conditions (Abbas et al. 2004). 

It should be stressed, however, that the reliable predictions for the alignment degree cannot be obtained with the "brute force approach". Indeed, radiative torques depend on many parameters, e.g. grain shape, grain size, radiation wavelength, grain composition, the angle between the radiation direction and the magnetic field. It is not practical to do numerical calculations for this vast multidimensional parameter space. Therefore the important empirical studies above demonstrated the radiative torque effects sometimes using one grain shape, one grain size, one wavelength of light, and one direction of the light beam in respect to magnetic field. The quantitative stage of radiative torque studies required theoretical models describing radiative torques. In Lazarian \& Hoang (2007a) we proposed a simple model of radiative torques which allowed a good analytical description of the radiative torque alignment.  
 
\subsection{Analytical Model}
\label{s:model}

\begin{figure}[h]
\includegraphics[width=3.3in]{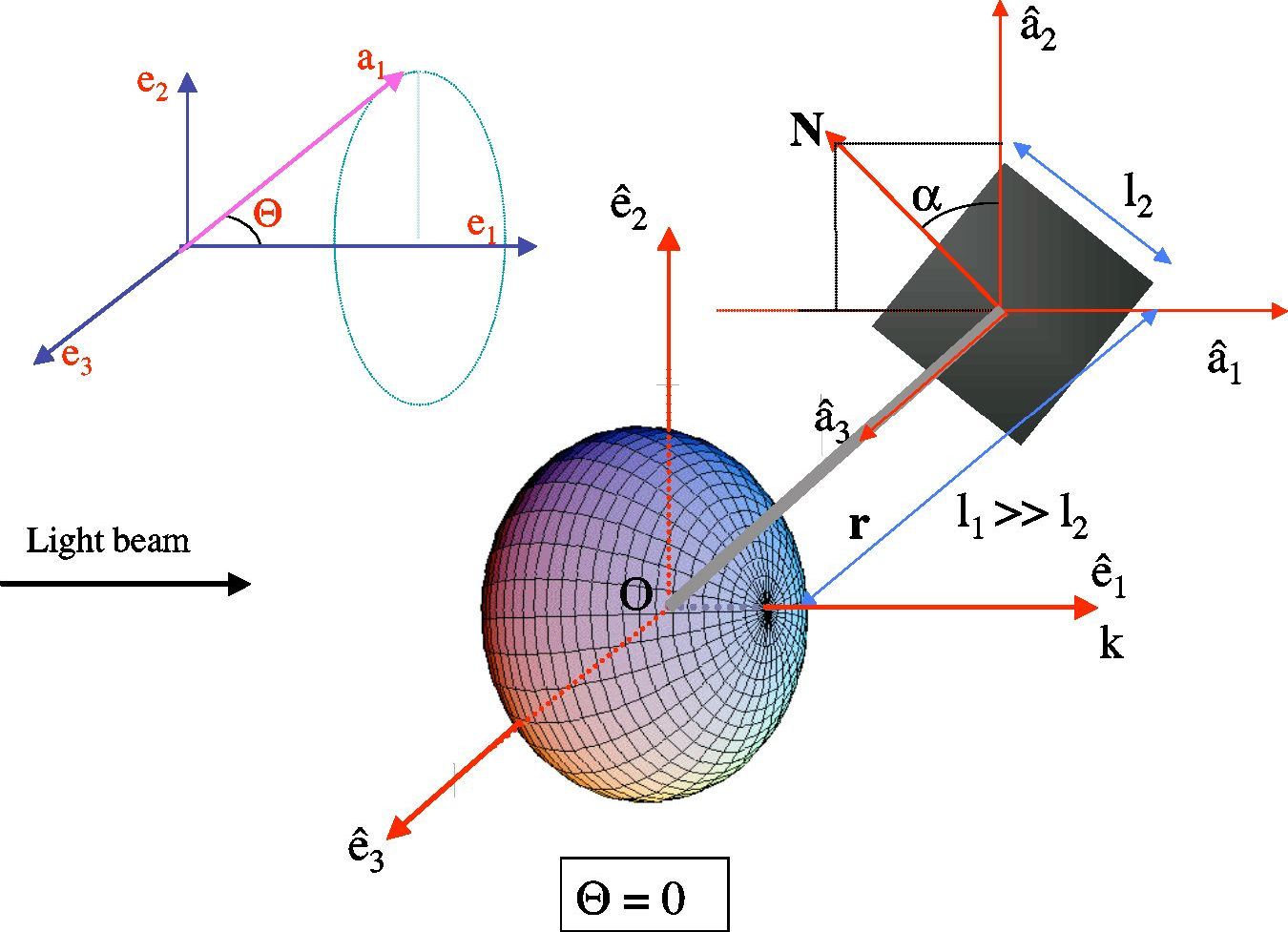}
\hfill
\includegraphics[width=2.5in]{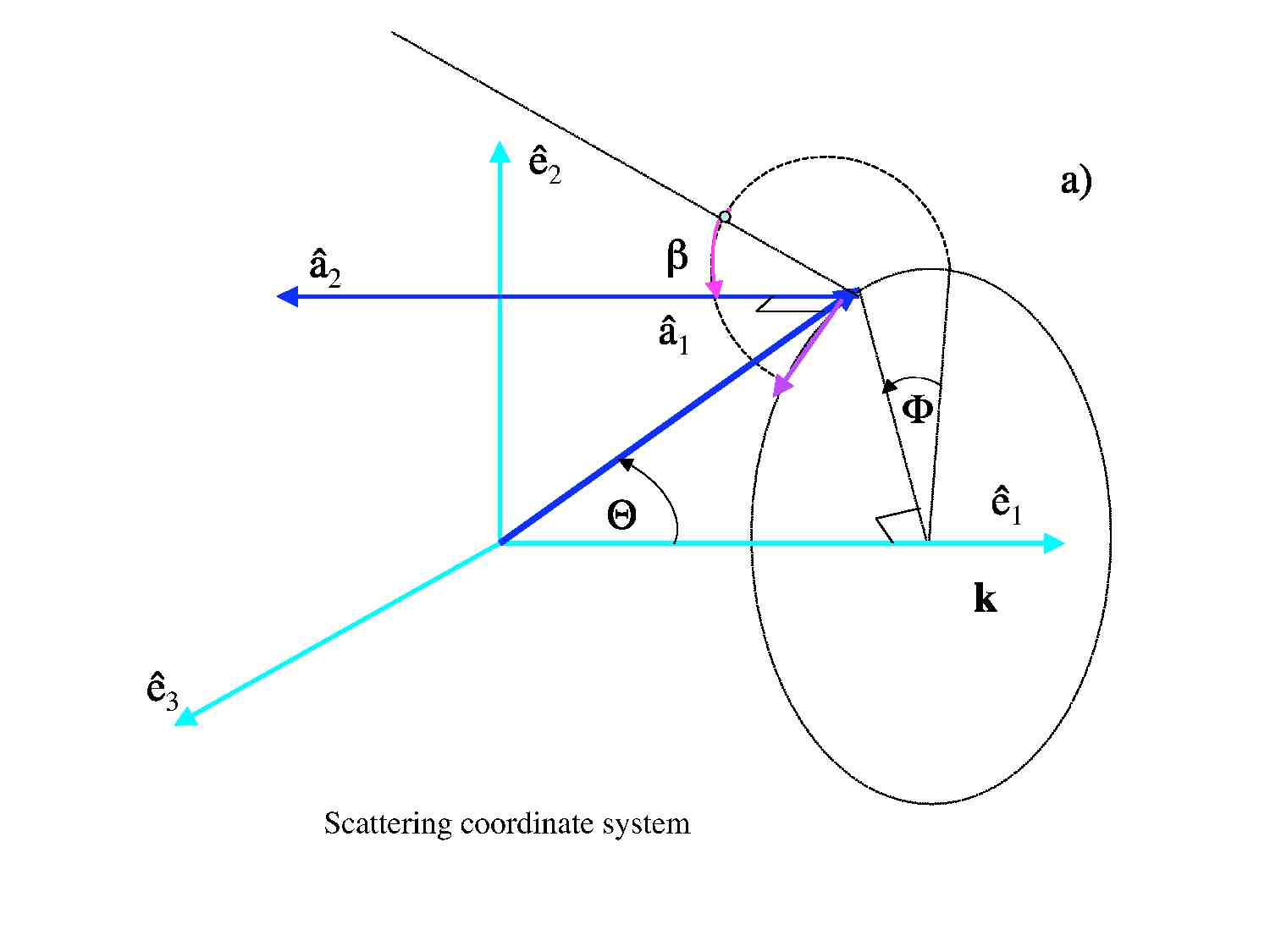}
\caption{
\small
{\it (a) Left panel}.-- A model of a ``helical'' grain,
that consists of a spheroidal grain with an inclined mirror attached to it,
reproduces well the radiative torques. From Lazarian \& Hoang 2007a.
 {\it (b) Right panel}.-- The ``scattering coordinate system'' which
illustrates the definition of torque components: ${\bf a}_1$ is directed
along the maximal inertia axis of the grain; ${\bf k}$ is the direction of radiation.
The projections of normalized radiative torques $Q_{e1}$,
$Q_{e2}$ and $Q_{e3}$ are calculated in this reference frame for $\Phi=0$.}
\label{AMO}
\end{figure}

Lazarian \& Hoang (2007a) showed that a simple model in Fig.~\ref{AMO} reproduces well the essential basic properties of radiative
torques. The model consists of an ellipsoidal grain with a mirror attached to
its side. Note,  that grains can be both  
``left-handed'' and  "right-handed". For our grain model to become ``right handed'' the mirror should be turned  by 90 degrees. 
Our studies with DDSCAT confirmed that actual irregular grains also vary in handedness. 

\begin{figure}
\includegraphics[width=2.9in]{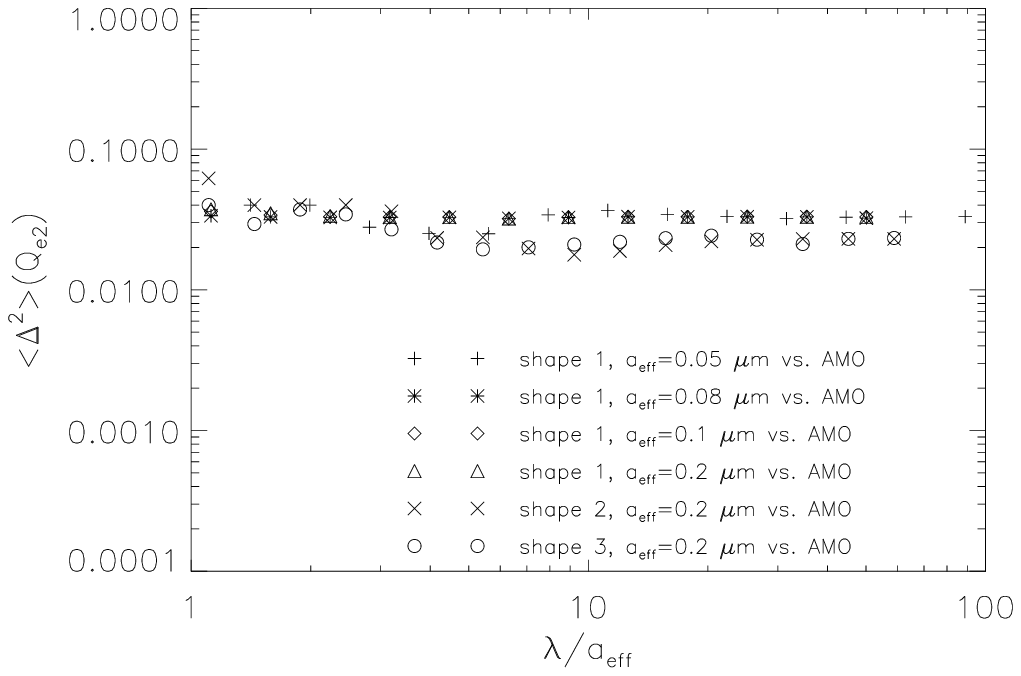}
\hfill
\includegraphics[width=2.9in]{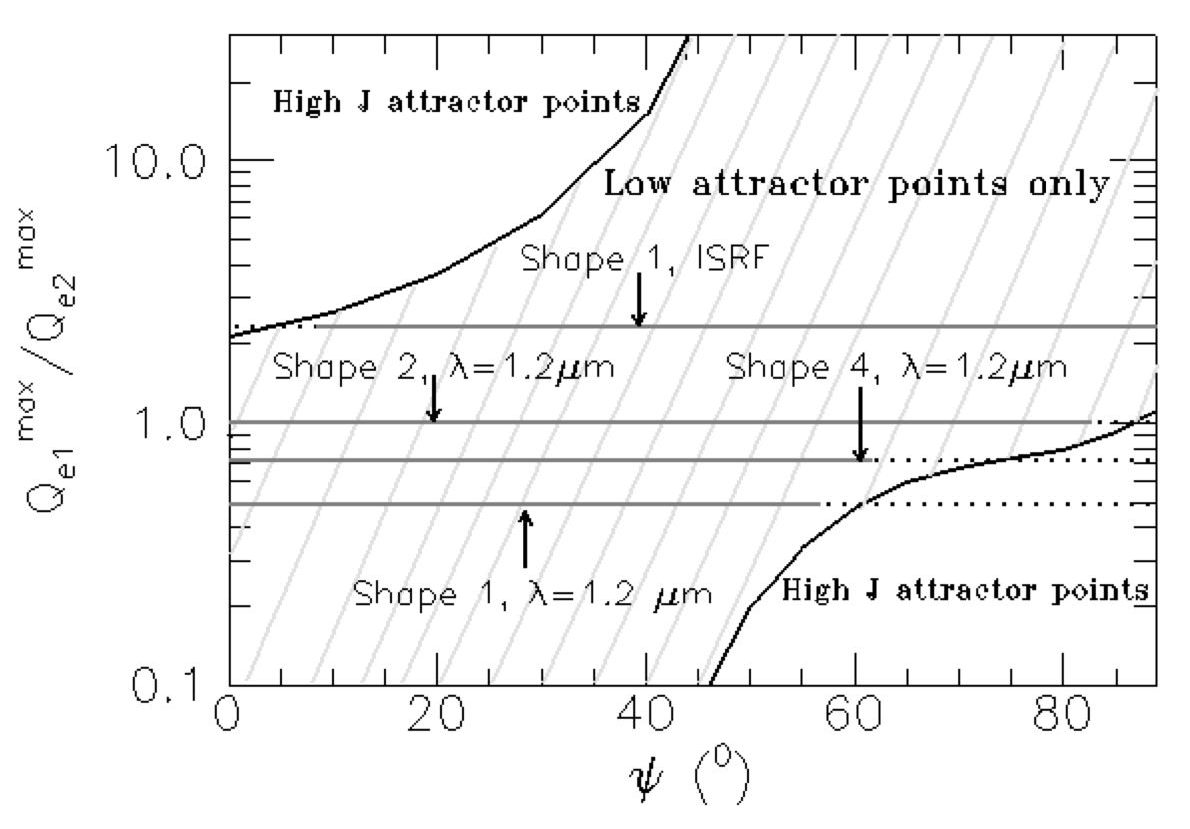}
\caption{\small {\it (a) Left panel}: Numerical comparison of the torques calculated with DDSCAT for irregular grains for different wavelength and the analytical model (AMO) of a helical grain. {\it (b) Right panel}: Parameter space for which grains have only low-$J$ attractor point and both low-$J$ and high attractor point. In the situation when the high-$J$ attractor point
is present grains eventually get there and demonstrate perfect alignment. In the situation when only low-$J$ attractor point is present, the alignment is partial. From Lazarian \& Hoang 2007a. }
\label{chi}
\end{figure}

The torques obtained analytically in the assumption of geometric optics for the model in Fig.~\ref{AMO} were shown to be in good agreement with the torques calculated
numerically with DDSCAT for irregular grains. To make a proper comparison Lazarian \& Hoang (2007a) choose a system of reference with the direction of  light along as vector ${\bf e_1}$,
and the grain axis of maximal inertia ${\bf a}_1$ being in the ${\bf e_1}$, ${\bf e_2}$ plane. As a result, we showed that for the problem of alignment only torques $Q_{e1}$ and
$Q_{e2}$ mattered. The third component $Q_{e3}$ happen to induce grain precession only, which for most situation is subdominant to the Larmor precession of the grain in the interstellar magnetic
field (see Table~1 in Lazarian 2007 for different time scales involved). The functional dependences of torques $Q_{e1}(\Theta)$ and $Q_{e2}(\Theta)$, where $\Theta$ is an angle between the
axis of maximal moment of inertia and the radiation direction, were shown to be very similar for the analytical model in Fig.~\ref{AMO} and irregular grains subject to radiation of different wavelengths. This remarkable
correspondence is illustrated in Fig.~\ref{chi}a using a function:
\begin{equation}
\langle \Delta^2\rangle(Q_{e2})=\frac{1}{\pi (Q_{e2}^{max})^2} \int^{\pi}_{0} \left[Q_{e2}^{irregular}(\Theta) -Q_{e2}^{model} (\Theta)\right]^2 d\Theta,
\label{chi_eq}
\end{equation}
which characterizes the deviation of the torques $Q_{e2}$ calculated numerically for irregular grains from the analytical prediction in Lazarian \& Hoang (2007a) model. 

While the functional dependence of torque components $Q_{e1}(\Theta)$ and $Q_{e2}(\Theta)$ coincides for grains of various shapes, their amplitudes vary for different  grains and different radiation wavelengths. In fact, Lazarian \& Hoang (2007a) showed that the radiative torque alignment can be fully determined if  
the ratio $q=Q_{e1}^{max}/Q_{e2}^{max}$ is known.  In terms of practical calculations, this enormously simplifies the calculations
of radiative torques: instead of calculating two {\it functions} $Q_{e1}(\Theta)$ and $Q_{e2}(\Theta)$ it
is enough to calculate just two {\it values} $Q_{e1}^{max}$ and $Q_{e2}^{max}$. According to Lazarian \& Hoang (2007a) the maximal value of the function $Q_{e1}(\Theta)$ is achieved for $\Theta=0$ of the function $Q_{e2}(\Theta)$ is achieved at $\Theta=\pi/4$. In other words, one can use a {\it single number} $q^{max}=Q_{e1}^{max}/Q_{e2}^{max}=Q_{e1}(0)/Q_{e2}(\pi/4)$ instead of
{\it two functions} to characterize grain alignment. Thus, it is possible to claim that the $q$-ratio is as important for the alignment as the grain axis
ratio for producing polarized radiation by aligned grains.

\subsection{Why do grains get aligned with long axes perpendicular to ${\bf B}$?}
\label{s:why}

Observations testify that interstellar grains tend to get aligned with long axes perpendicular to the ambient magnetic field, the fact that was frequently used to argue that the Davis-Greenstein (1951) mechanism is responsible for the alignment. Can radiative torques explain this observational fact?
\begin{figure}[h]
\epsfig{file=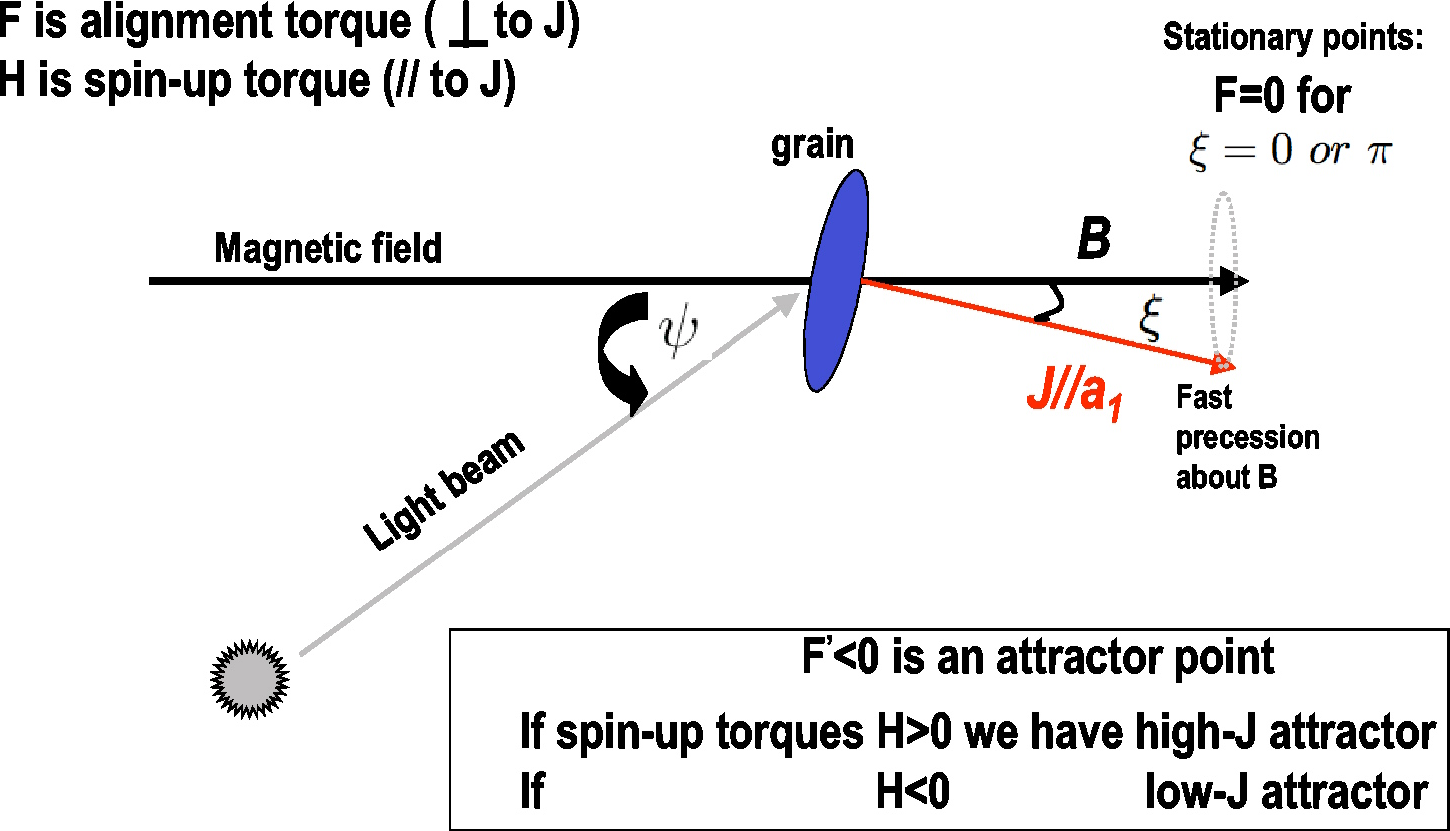, width=7.73cm, angle=270}
\caption{A simplified explanation of the grain alignment by radiative torques.The grain, which is depicted as a spheroid in the figure, in fact, should be irregular to get non-zero radiative torque. 
The positions ${\bf J}$ parallel (or anti-parallel) to ${\bf B}$ correspond to the stationary points as at these positions the component of torques that changes the alignment angle vanishes. As 
internal relaxation makes ${\bf J}$ aligned with the axis $a_1$ of the maximal moment of grain inertia, the grain gets aligned with long axes perpendicular to ${\bf B}$.}
\label{alignment}
\end{figure}

Fig.~\ref{alignment} illustrates why radiative torques tend to align grains the "right way", i.e. in agreement with observations. Interstellar grains experience internal relaxation that tends to make them rotate about their axis of maximal moment of inertia. Therefore, it is sufficient to follow the dynamics of angular momentum to determine grain axes alignment. Let us call the component of torque parallel to ${\bf J}$ the {\it spin-up torque}
${\bf H}$ and perpendicular to ${\bf J}$ the {\it alignment torque} ${\bf F}$. The angular momentum ${\bf J}$ is precessing about magnetic field due to the magnetic moment of a grain (see
Dolginov \& Mytrophanov 1976). The alignment torques ${\bf F}$ are perpendicular to ${\bf J}$ and therefore as  ${\bf J}$ gets parallel to ${\bf B}$ the fast precession of the grain makes the torques averaged over ${\bf J}$ precession vanish as $\xi\rightarrow 0$. Thus the positions corresponding to ${\bf J}$ aligned with ${\bf B}$ are stationary points, irrespectively of the functional forms of radiative torques, i.e.
of components $Q_{e1}(\Theta)$ and $Q_{e2}(\Theta)$. In other words, grain can stay aligned with $\xi=0$ or $\pi$.

The arguments above are quite general, but they do not address the question whether there are other stationary points, e.g. whether the alignment can also happen with ${\bf J}$ perpendicular to ${\bf B}$.
To answer this question one should use the actual expressions for  $Q_{e1}(\Theta)$ and $Q_{e2}(\Theta)$. The analysis in Lazarian \& Hoang (2007a) shows that there is, indeed, a range of angles
between the direction of radiation and the magnetic field for which grains tend to aligned in a "wrong" way, i.e. with long axes parallel to magnetic field. However, this range of angles is rather narrow and does not exceed several degrees. Moreover, the "wrong" alignment corresponds to the positions for which the spin-up torques ${\bf H}$ are negative, which induces grain alignment with low angular momentum\footnote{Within this short report we do not go into the rather complex details of grain thermal wobbling which stabilize the value of $J$ at some fraction of the thermal value. The corresponding discussion is presented in Lazarian \& Hoang (2007a) and Hoang \& Lazarian (2008a).}. Grain thermal wobbling at low-$J$ attractor point (Lazarian 1994, Lazarian \& Roberge 1997) induces the variations in the angle that is typically exceeds this range. Thus a remarkable fact emerges: grains get always aligned with long axes perpendicular to ${\bf B}$!

\subsection{Quantitative predictions available}
\label{predictions}

For typical interstellar conditions radiative torques absolutely dominate other torques that can induce alignment. Therefore the alignment problem gets  independent of the magnitude
of radiative torques and can be fully described for a given grain by defining a single parameter $q^{max}$ (see \S\ref{s:model}).  
The practical use of the $q$-ratio is presented in Fig.~\ref{chi}b where the parameter space for the existence of
only low-$J$ as well as  the coexistence of both high-$J$ and low-$J$ attractor points is shown. In the case when high-$J$ attractor point exists, grains eventually get into that position, which is
characterized by both the fast rotation and {\it perfect} alignment in respect to magnetic field. In the situation when
only low-$J$ attractor point exists, the alignment in respect to magnetic field is not stable and the characteristic degrees
of alignment vary on average from 20\% to 50\%. We see from Fig.~\ref{chi}b that when the grain alignment is dominated by radiative torques 
for a substantial part of the parameter space grains are driven to the low-$J$ states, i.e. {\it subthermally}, which is in contrast to
the assumption in Draine \& Weingartner (1996) that in the presence of radiative torques
most of the interstellar grains should rotate at  $T_{rot}\gg T_{gas}$.
\begin{figure}
\includegraphics[width=2.8in]{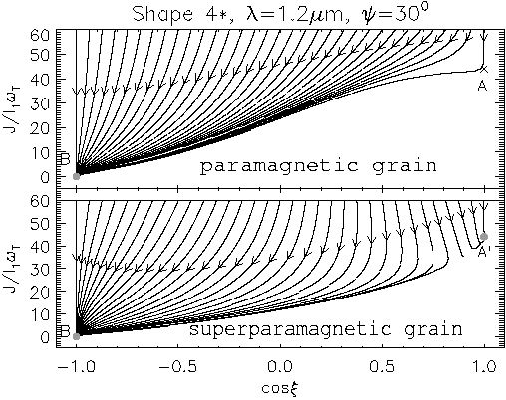}
\hfill
\includegraphics[width=2.7in]{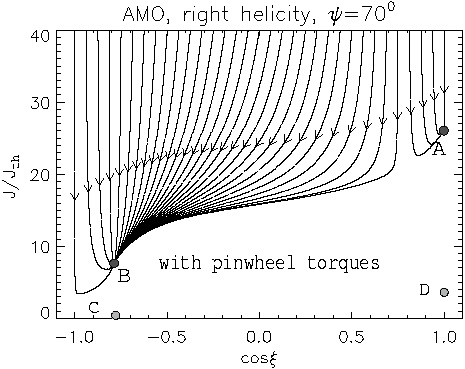}
\caption{\small Phase trajectories in terms of $J/J_{th}$, where $J_{th}$ is the thermal angular momentum $I\omega_{T}$, and the alignment angle $\xi$. {\it (a) Left Panel}: A paramagnetic grain gets only a low-$J$ attractor point. For the same set of parameters a superparamagnetic grain gets also a high-$J$ attractor point.
The fact that most of the phase trajectories go in the direction of the low-$J$ attractor point illustrates the dominance of the radiative torques for the alignment even
in the case of the superparamagnetic grain. However, high-$J$ attractor points are more stable than the low-$J$ attractor points. As a result, all grains eventually end up
at the high-$J$ attractor point. {\it (b) Right Panel}:  Grain alignment by radiative torques in the presence
of  pinwheel torques. The shown case corresponds to the presence of both the low-$J$ and high-$J$ attractor points in the absence of pinwheel torques. In the case when only a low-$J$ attractor
point exists the strong pinwheel torques lift the low-$J$ attractor point enhancing the alignment. From Lazarian \& Hoang 2008 and Hoang \& Lazarian 2008b.}
\label{superparamagnetic}
\end{figure}  

Superparamagnetic grains, i.e. grains with enhanced paramagnetic relaxation, 
 were invoked by Jones \& Spitzer (1967) within the model of paramagnetic alignment (see also arguments in support of the existence of such grains in Bradley 1994, Martin 1995, Goodman \& Whittet 1995).
   What does happen with superparamagnetic grains subject to radiative torques.
  Lazarian \& Hoang (2008)  found such grains {\it always} get 
a high-$J$ attractor point. 

To demonstrate the radiative torque alignment we use phase diagrams (see Fig.~\ref{superparamagnetic})
where large angular momentum is given for grains initially at different alignment angle $\xi$ and sufficiently large $J$ (see Fig~\ref{alignment}). 
Fig.~\ref{superparamagnetic}a shows that for superparamagnetic grains subject to a diffuse interstellar radiation field most grains still get
to the low-$J$ attractor point, which reflects the fact that it is the radiative torques that dominate the
alignment. As the high-$J$ attractor point is more stable compared to the low-$J$ attractor point, similar to the ordinary paramagnetic  grains,
superparamagnetic grains get transfered by gaseous collisions from the low-$J$ to high-$J$ attractor points. Thus, superparamagnetic grains 
always rotate at high rate in the presence of radiative torques. One concludes that, rather
unexpectedly, intensive paramagnetic relaxation changes the rotational state of the grains, enabling them to
rotate {\it rapidly}. The alignment of grains at high-$J$ point is {\it perfect}.

Pinwheel torques were considered by Purcell (1979) in the context of paramagnetic alignment.
How do these torques also affect the radiative torque alignment? Hoang \& Lazarian (2008b) showed
that the sufficiently strong pinwheel torques can create new high-$J$ attractor points (see Fig.~\ref{superparamagnetic}b) Therefore for
sufficiently strong pinwheel torques, e.g. for torques arising from H$_2$ formation, one may observe the correlation
of higher degree of polarization with the atomic hydrogen content in the media, provided that H$_2$ torques as strong
as they considered in Purcell (1979) and the subsequent papers (see Spitzer \& McGlynn 1979, Lazarian 1995, Lazarian \& Draine 1997). 
Interestingly enough, just detection of the variations of grain alignment with the expected magnitude of pinwheel torques will discard 
the superparamagnetic hypothesis. Indeed, for superparamagnetic grains the alignment should be perfect irrespectively of other factors (see above).

\subsection{Synergy of CMB polarization and grain alignment studies}

Understanding of grain alignment is essential for the successful separating of the CMB polarization from
the polarization arising from dust. Any construction of a polarization template for the dust foreground uses 
assumptions on grain alignment. Quantitative predictions of grain alignment (see \S\ref{predictions}) allow quantitative testing.

The progress in understanding of grain alignment can be gauged by modeling of polarization in interstellar gas, molecular clouds,
accretion disks. The correspondence of the theory and observations obtained along these directions is encouraging (see Lazarian 2007).
However, more detailed modeling is required. We hope that by the time of CMBPol mission the grain alignment theory will be better tested
and therefore trustworthy. However, additional  and sometimes unique tests, e.g. at 30-100~GHz, that CMBPol data can provide will be extremely
useful. To test theoretical predictions (see \S\ref{predictions})
a larger coverage of frequencies is advantageous. 

Grain alignment can place stringent constraints on the abundance of strongly magnetic, e.g. "supermagnetic" grains. Such
grains can be a source of an additional emission magnetodipole emission in the range of 30-90~GHz (Draine \& Lazarian 1999). Our work also suggestive
that large Zodiacal dust grain particles are likely to be aligned and therefore present an additional polarized foreground. In the report above  we did not talk 
about possible processes of alignment of PAH particles which are responsible for the "spinning dust" emission (Draine \& Lazarian 1998ab). The only process
that has been considered for them so far is the paramagnetic resonance relaxation alignment introduced in Lazarian \& Draine (2000). Other processes
should be explored, however.\\ 
  
{\bf Acknowledgments}. I acknowledge the support by the NSF grant  AST-0507164,
as well as by the NSF Center for Magnetic Self-Organization in Laboratory and Astrophysical
Plasmas.

%-----------------------------------------------------------------------------------------------
% Section - Bibliography
%-----------------------------------------------------------------------------------------------

\newpage
\addcontentsline{toc}{section}{References} 
\bibliography{pdebiblio}
\bibliographystyle{hapj}

\end{document}